# Observing *Zitterbewegung* for photons near the Dirac point of a two-dimensional photonic crystal

**Xiangdong Zhang**
Department of Physics, Beijing Normal University, Beijing 100875, China

## Abstract

It is shown, for the first time, that the *zitterbewegung* of photon can appear near the Dirac point in two-dimensional photonic crystal. The superiority of such a phenomenon for photons is that it can be found in different scaling structures with wide frequency regions. It can be observed by measuring the time dependence of the transmission coefficient through photonic crystal slabs. Thus, it is particularly suited for experimentally observing this effect. We have observed such a phenomenon by exact numerical simulations, confirming a long-standing theoretical prediction.

Since the original work of Schroedinger [1], the notion of *Zitterbewegung* (trembling motion) has been a long-standing theoretical prediction of relativistic quantum mechanics. *Zitterbewegung* (ZB) represents an oscillatory motion of free electrons described by the Dirac equation in the absence of external fields, which is caused by the interference between the positive and negative energy states. The characteristic frequency of this motion is determined by the gap between the two states, which is of order $10^{20}$ Hz. It was believed that the experimental observation of the effect is impossible since one would confine the electron to a scale of the Compton wavelength [2].

Due to a formal similarity between two interacting bands in a solid and the Dirac equation for relativistic electrons in a vacuum, theoretical investigations have shown that the ZB can also appear even for a nonrelativistic particle moving in a crystal [3, 4], or for quasiparticles governed by the Bogoliubov–de Gennes equations in superconductors [5], or for spintronics in some semiconductor nanostructures with spin-orbit coupling [6-9]. Recently, graphene has become the subject of intensive research due to the successful fabrication experimented by Novoselov et al. [10]. Graphene is a monolayer of carbon atoms densely packed in a honeycomb lattice, which can be viewed as either an individual atomic plane pulled out of bulk graphite or unrolled single-wall carbon nanotubes [11,12]. Its electronic structure bears remarkable analogies with the solutions of the Dirac equation. Thus, Zitterbewegung-like phenomena in graphene and carbon nanotubes have been discussed within the last two years [13-15]. A unified description of ZB for spintronic, graphene, and superconducting systems has also been given [16]. However, the above investigations on the ZB are for electrons in solid systems. Because of the complex interactions in these systems, it is very difficult to observe it directly from experimental measurements.

Analogous to the above electron systems, in this work we will demonstrate that the ZB can also appear for photon transport in two-dimensional (2D) photonic crystal (PC). The experimental observation of ZB in the electron case is severely hindered by the complex interactions and the limited scale. No such difficulties exist in the PC. Thus, this makes it possible to observe such a phenomenon directly.

We consider a 2D triangular lattice of cylinders immersed in air background with lattice constant $a$. The radius ($R$) and the dielectric constant ($\varepsilon$) of the cylinders are $0.3a$ and 11.4, respectively. The band structure of the system for P wave obtained by the multiple-scattering

Korringa–Kohn–Rostoker method is shown in Fig. 1. The key feature of this band structure is that the band gap may become vanishingly small at corners of the Brillouin zone at $f = 0.466\ \omega a / 2\pi c$, where two bands touch as a pair of cones in a linear fashion (bottom in right panels of Fig.1). Such a conical singularity is also referred to as the Dirac point ($\omega_D$). According to the analyses of Refs. [17, 18], photon transport near the Dirac point in the PC can also be described by the following Dirac equation

$$H \begin{pmatrix} \psi_1 \\ \psi_2 \end{pmatrix} = \delta\omega \begin{pmatrix} \psi_1 \\ \psi_2 \end{pmatrix}, \quad \delta\omega = \omega - \omega_0 \tag{1}$$

with

$$H = \begin{pmatrix} 0 & -iv_D(\partial_x - i\partial_y) \\ -iv_D(\partial_x + i\partial_y) & 0 \end{pmatrix}. \tag{2}$$

Where $\psi_1, \psi_2$ represents the amplitudes of two degenerate Bloch states at one of the corners of the hexagonal first Brillouin zone (top in right panels of Fig.1). The frequency $\omega_D$ and velocity $v_D$ in the Dirac point depend on the structure of the PC. If such an analysis is correct, the conductance of photons around the Dirac point should be inversely proportional to the thickness of the sample, which is similar to diffusion behavior of waves through a disordered medium [18]. Recently, such a phenomenon has been demonstrated by exact numerical simulations [19]. This means that photon transport near the Dirac point in the PC can actually be described by the Dirac equation. Then, it is natural to ask whether or not some basic phenomena originating from the Dirac equation such as ZB can be also observed in the PC. In the following, we will explore the possibility of observing such a long-standing theoretical prediction in the PC systems.

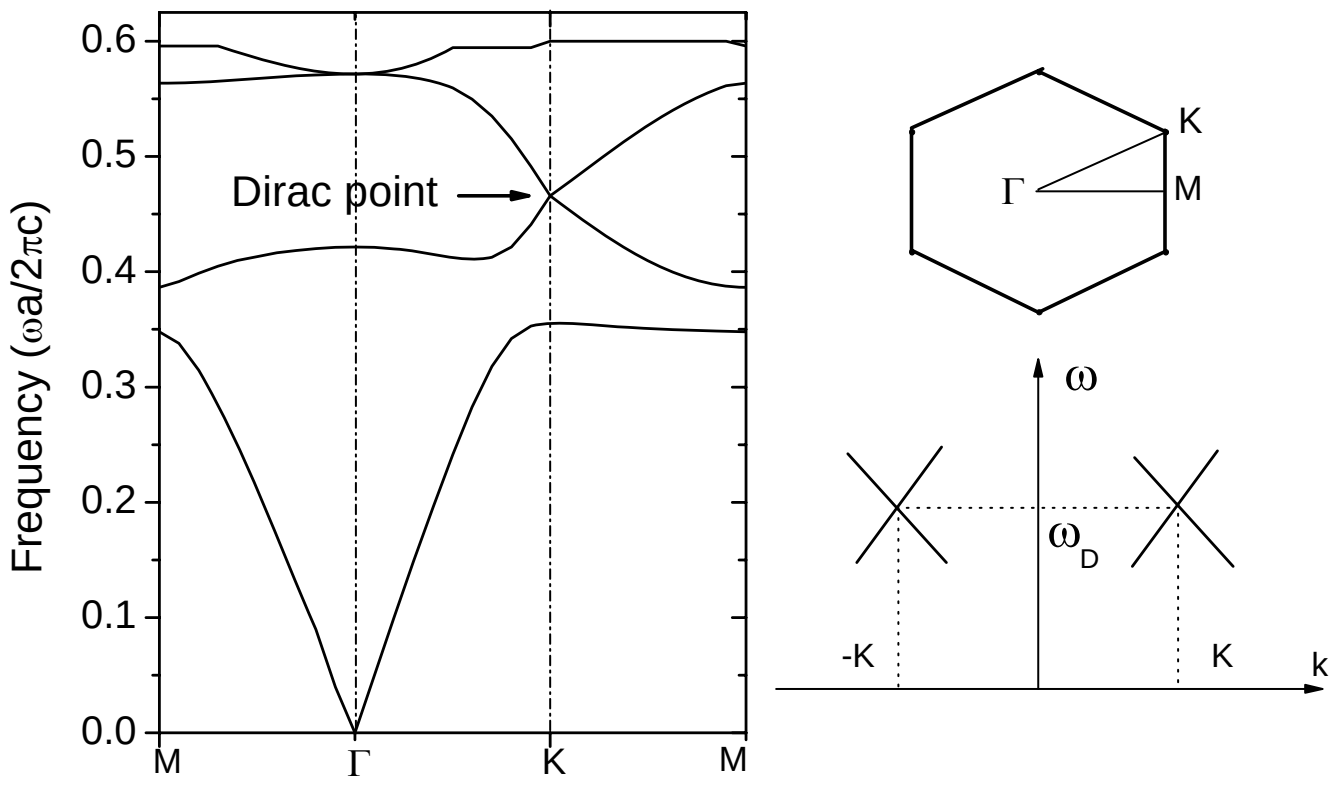

**Fig.1 Left panel: Calculated photonic band structure of P wave for a triangular lattice of dielectric cylinder with R/a=0.3 and $\varepsilon = 11.4$ in an air background. Right panels: Hexagonal first Brillouin zone of the PC (top) and the model for linear dispersion relation of the doublet near one of the zone corners (bottom).**

Thus, we consider the dynamic behavior of photon transport in the above PC near the Dirac point. The time dependence position operator in the Heisenberg picture is written as

$$\vec{r}(t) = e^{iHt/\hbar}\vec{r}(0)e^{-iHt/\hbar}. \qquad (3)$$

Inserting Eq.(2) in Eq.(3), explicit results for two coordinate components of $\vec{r}(t)$ can be obtained as

$$\begin{aligned} x(t) &= x_0 + v_D\sigma_x t + \frac{k_y}{2k^2}\sigma_z[1-\cos(2v_D kt)] \\ &+ \frac{k_y}{2k^3}(k_x\sigma_y - k_y\sigma_x)[2v_D kt - \sin(2v_D kt)] \end{aligned}, \qquad (4)$$

$$\begin{aligned} y(t) &= y_0 + v_D\sigma_y t - \frac{k_y}{2k^2}\sigma_z[1-\cos(2v_D kt)] \\ &- \frac{k_y}{2k^3}(k_x\sigma_y - k_y\sigma_x)[2v_D kt - \sin(2v_D kt)] \end{aligned}. \qquad (5)$$

Here the coordinate component $x(t)$ is taken along the $\Gamma K$ direction of the PC and $y(t)$ along the $\Gamma M$ direction. For a Gaussian wave packet, the calculations are also straightforward. We consider a Gaussian wave packet in the form [6, 14]

$$\psi(\vec{r},0) = \frac{1}{2\pi}\frac{d}{\sqrt{\pi}}\int d^2\vec{k}e^{-\frac{1}{2}d^2k_x^2-\frac{1}{2}d^2(k_y-k_{0y})^2}e^{i\vec{k}\cdot\vec{r}}\begin{pmatrix}1\\0\end{pmatrix}. \qquad (6)$$

The packet is centered at $k_0 = (0, k_{0y})$ and is characterized by a width of d. The unit vector (1, 0) is a convenient choice [6, 14]. Selecting the [11] component of $x(t)$, a direct calculation gives the average of $x_{11}(t)$ as

$$\begin{aligned} \overline{x}(t) &= <\psi(\vec{r},0)|x_{11}(t)|\psi(\vec{r},0)> \\ &= \frac{d^2}{\pi}\int_{-\infty}^{\infty}\int_{-\infty}^{\infty}\frac{k_y}{2k^2}[1-\cos(2v_D kt)]e^{-d^2k_x^2-d^2(k_y-k_{0y})^2}dk_x dk_y \end{aligned}. \qquad (7)$$

After the average displacement is obtained, the velocity of the photons can be obtained by $\overline{v}(t) = \partial\overline{x}(t)/\partial t$. The calculated results for $\overline{x}(t)$ and $\overline{v}(t)$ as functions of time are plotted in Fig.2. Figure 2 (a) and (b) represent the results for different packet widths d at $k_{0y} = 0.47/a$ and $v_D \approx c/3$, while (c) and (d) are those for different $k_{0y}$ at $d = 30a$. Here c is the velocity of the photons in a vacuum. It can be seen clearly that all of them are oscillations, which is a direct manifestation of ZB. The period of ZB depends weakly on d, while it is linearly dependent on $k_{0y}$ or $\delta\omega$ due to linear Dirac-like dispersion around the Dirac point. For example, the periods for $\bar{x}(t)$ and $\bar{v}(t)$ at $k_{0y} = 0.05/a$ are 4 times of those at $k_{0y} = 0.2/a$ (see Fig.2 (d) and (c)). In contrast, the amplitudes of ZB depend strongly on d. For small d there are almost no oscillations, for very large d the ZB oscillations are nearly undamped. In addition, the ZB for all cases has a transient character, as it is attenuated exponentially with the increase of time. This can be understood from Eq.(7), in which there are terms for exponential decay. These features are similar to the previous analytical results for the electrons in graphene [13-15].

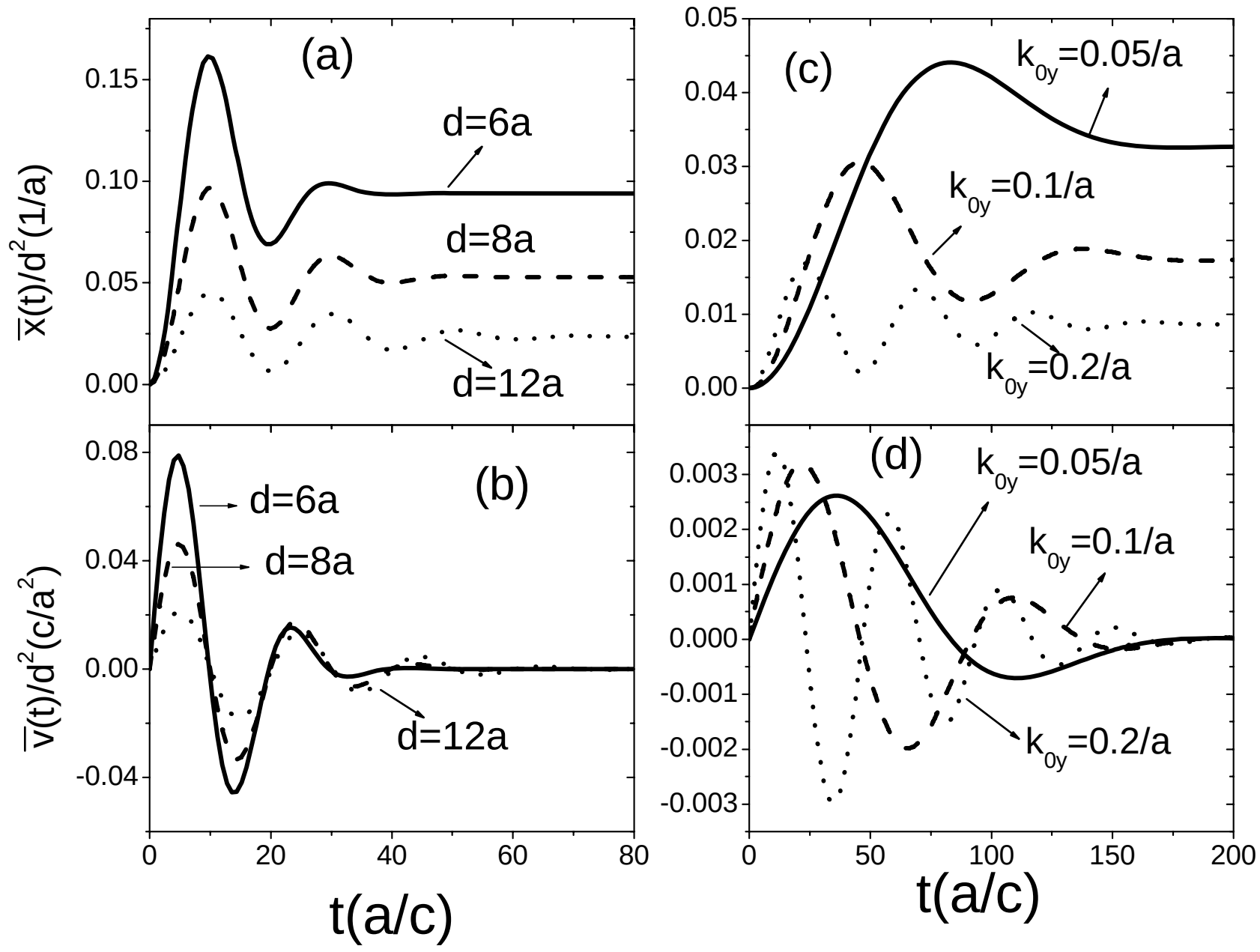

**Fig.2 (a) and (b) correspond to Zitterbewegung of displacement and velocity of photon in PC as a function of time for a Gaussian wave packet with different packet width d at $k_{0y} = 0.47/a$ and $v_D = c/3$. The values of d are marked in the figures. (c) and (d) are the corresponding results with different $k_{0y}$ at d=30a. The values of $k_{0y}$ are also marked in the figures. The parameters of PC are identical to those in Fig.1.**

However, in contrast to the electron cases, the present results satisfy the scaling relation, because the structure parameters of the PC are not confined to certain values. That is to say, the above PC structures are composed of pure dielectric materials and can be scaled to different frequency regions, from microwave to optical frequency. As a result, the ZB for photons in the PC can be observed within a wide frequency region. In addition, the complex interactions in the electron systems do not exist in the present PC system. This facilitates observation of ZB in the PC. In fact, the mean photon number flux $j(t)$ is proportional to the velocity of photon $\overline{v}(t)$ ( $j(t) \propto \overline{v}(t)$ ), and $j(t)$ can be obtained by making a difference of pulse intensity (I(t)) to time ( $j(t) = dI(t)/dt$ ) [21, 18]. Thus, we can observe the ZB of photons by measuring the intensity of optical pulse as a function of time.

In order to observe such a phenomenon, we take above PC sample with width $W$ and thickness $L$ . We inject a Gaussian pulse $(1/(\sqrt{2\pi}\Delta f)\exp[-(f - f_c)^2/(2\Delta f^2)])$ into it and calculate the time dependence of intensity inside or outside the sample by using the exact multiple-scattering method [20]. The center frequency ( $f_c$ ) of the pulse is taken as $0.466\omega a/2\pi c$ . In all the calculations discussed below, $W$ is chosen to be $100\ a$ . The time dependence of the intensities at different positions inside the sample with L=32a and pulse width $\Delta f = 0.02\omega a/2\pi c$ are shown in Fig.3. The dark-solid line, dashed line and dotted line correspond to the results inside the sample at (x=0.0, y=0.5a), (x=8.0a, y=0.5a) and (x=16a, y=0.5a), respectively. Here the coordinate center is located at the center of the sample. The red-solid line is the result outside, at a distance of 8.0*a* from the right surface of the sample. Here all curves are normalized. The oscillation feature can be observed clearly and the periods of oscillation (time interval between two peaks) are the same for all curves. That is to say, the evolution of intensity outside the sample exhibits the same oscillation features to those inside the sample. This means that we can obtain evolution information of the pulse inside the sample by measuring the transmission coefficient (or intensity outside the sample) as a function of time.

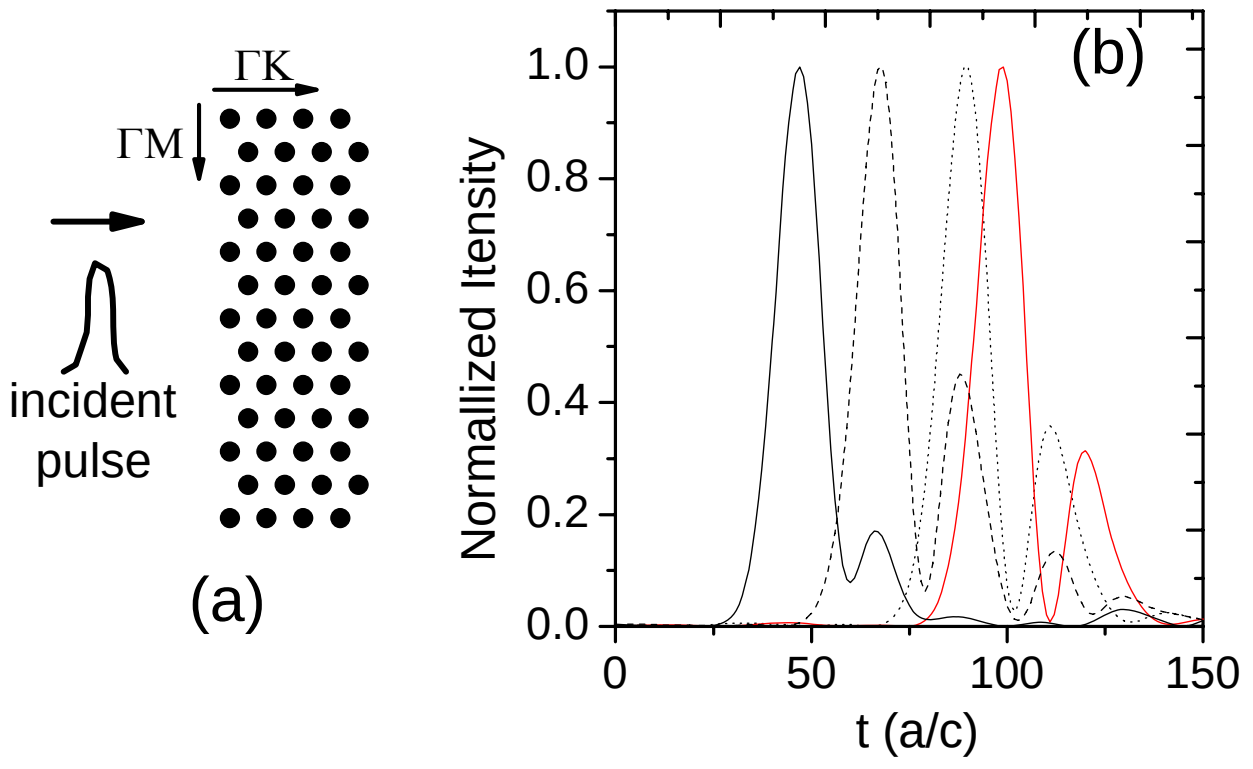

**Fig.3 (a) Schematic picture depicting simulation processes. (b) After the injection of a Gaussian pulse with center frequency $f_c = 0.466\omega a/2\pi c$ and pulse width $\Delta f = 0.02\omega a/2\pi c$ into the sample with L=32a, the time dependence of the intensity at different positions. Dark-solid line, dashed line and dotted line correspond to the results inside the sample at y=0.5a and x=0.0, 8.0a and 16.0a, respectively. Red line is the result outside the sample at a distance of 8.0a from the right surface. The other parameters are identical to those in Fig.1.**

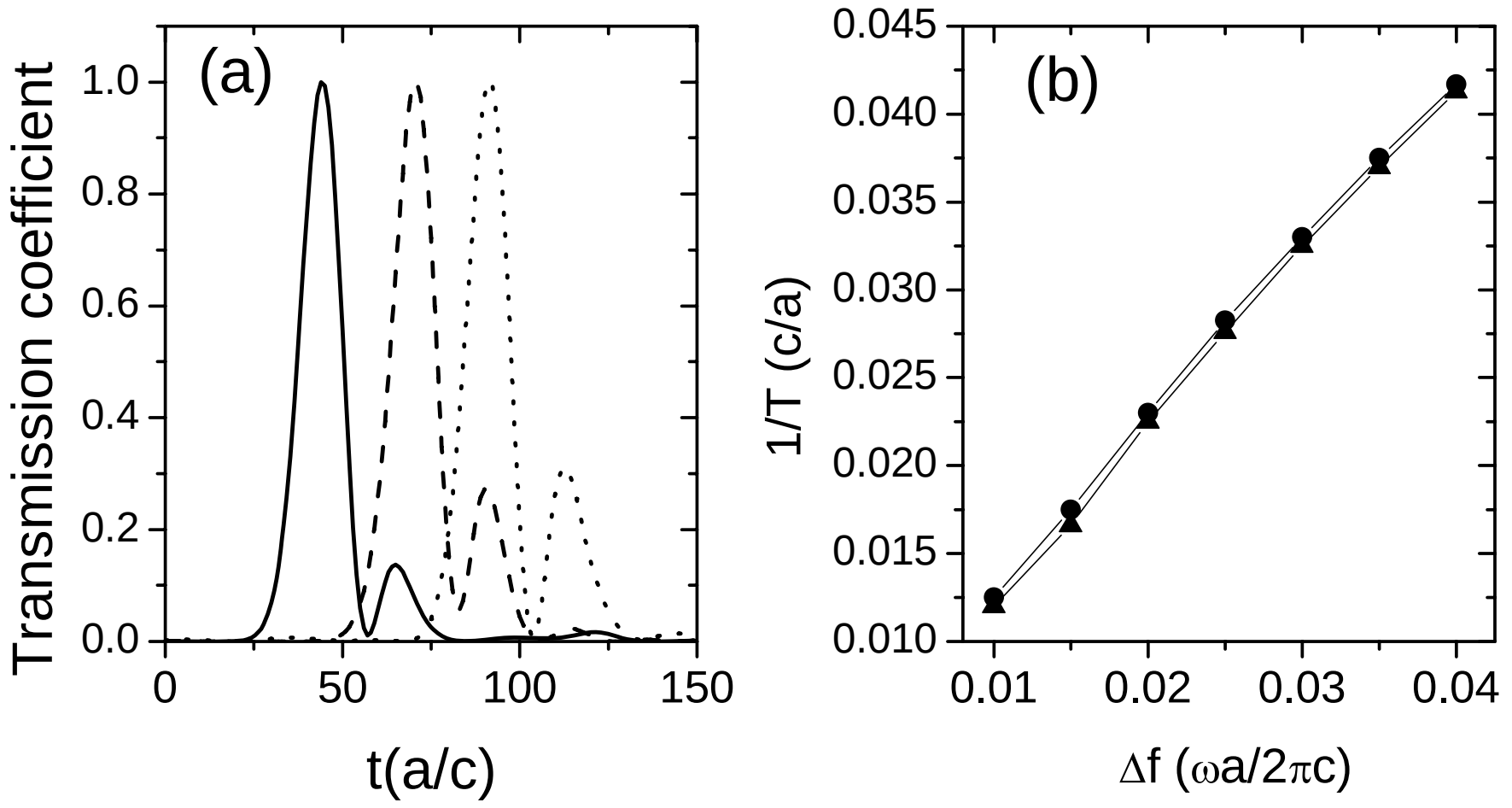

**Fig.4 (a) After the injection of a Gaussian pulse with center frequency $f_c = 0.466\omega a/2\pi c$, the time dependence of transmission coefficient for the samples with different thicknesses. Solid line, dashed line and dotted line correspond to the results with L=14a, 24a and 32a, respectively. (b) The inverse values of oscillation period (1/T) as a function of pulse width $\Delta f$. Circle and triangular dots correspond to the cases with L=14a, 32a, respectively. The other parameters are identical to those in Fig.3.**

The time dependence of the transmission coefficients in the present cases exhibit the following oscillation features. Firstly, the period of oscillation (T) depends weakly on the thickness of the sample. Figure 4 (a) shows the transmission coefficients as a function of time for the PC samples with different thicknesses. The solid line, dashed line and dotted line correspond to the results with L=14a, 24a and 32a, respectively. It is seen that the period of oscillation increases slightly with the increase of sample thickness. Secondly, the ratios of intensities between the second peak and the first peak increase considerably. For thin samples such as $L \leq 3a$, the second peak basically disappears . That is to say, no oscillation can be observed. This is because of the finite size effect of the sample. When the thickness of the sample is very small, wave transport exhibits ballistic behavior. In this case, it can not be described by Eq.(1). With the increase of sample thickness, diffusion behavior appears gradually. When the thickness of the sample is bigger than 10a, diffusion behavior plays a leading role. In such a case, remarkable oscillations can be observed. Thirdly, the period of oscillation is linearly dependent on the pulse width and is not determined by the central frequency. This can be seen clearly from Fig.4 (b). Circular and triangular markers in Fig.4 (b) are the simulation results of 1/T as a function of pulse width for the cases with L=14a, 32a, respectively. These features are completely different from the oscillations of the wave in a dielectric slab or cavity originating from the interface reflection or Fabry-Perot effect, in which the oscillation period is determined by the thickness of the sample and the central frequency of the pulse. In addition, the features are also not related to the localization effect, because wave transport exhibits diffusion behavior. Such a phenomenon is directly related to the ZB.

In fact, the pulse width $\Delta f$ in the simulations corresponds to $\delta\omega$ in Eq.(1). ($\delta\omega = \omega - \omega_0 = 2\pi\Delta f$ ). The linear dependence of T on $\Delta f$ in Fig.4 (b) agrees with the linear relation of ZB period with $\delta\omega$ in the theory of ZB. Furthermore, the oscillation periods of $j(t)$ obtained from exact numerical simulations are also identical with those of $\overline{v}(t)$ in Fig.2 (d). This can be seen more clearly from Fig.5.

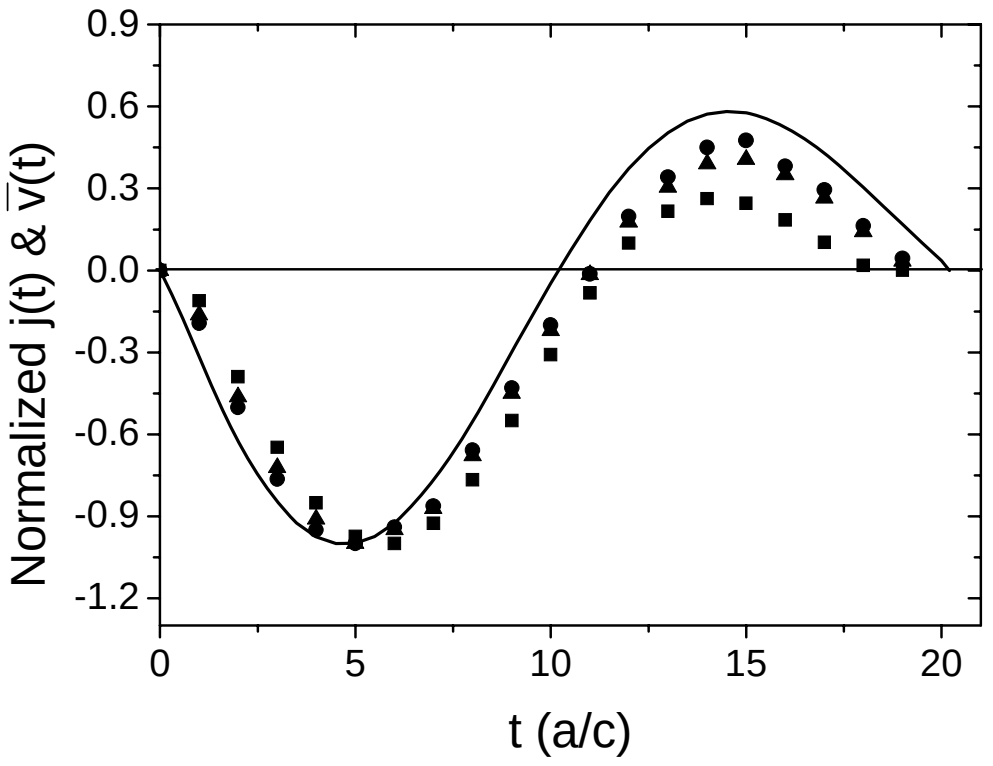

**Fig.5 Comparison between the theoretical result (solid line) for the velocity ($\overline{v}(t)$ ) of pulse evolution and numerical simulations for the mean photon number flux ( $j(t)$ ) (circle, triangular and square dots). Square, triangular and circle dots correspond to the cases with L=14a, 32a and 40a, respectively. Here the theoretical results are normalized and $\Delta f = 0.025\omega a / 2\pi c$ . The other parameters are identical to those in Fig.4.**

Figure 5 shows the comparison between the theoretical result for $\overline{v}(t)$ (solid line) and exact numerical simulations for $j(t)$ (circular, triangular and square markers). Here the center frequency and the pulse width are taken as $0.466\omega a / 2\pi c$ and $0.025\omega a / 2\pi c$ , respectively. The parameter $v_D = c / 2.7$ is obtained exactly by calculating the ratio between the transport distance of pulse and time. The square, triangular and circular markers represent the mean photon number fluxes for the cases with L=14a, 32a and 40a, respectively. They are obtained by making a difference of the transmission coefficients to time. Comparing them, we find that the oscillating periods are basically identical. Some difference between the theoretical results and numerical simulations originate from the effect of finite size and interface. With the increase of sample thickness, the simulation results gradually agree more with the theoretical estimation. These indicate that the oscillations of the transmission coefficients or the intensities in Fig.3 and 4 are actually a direct manifestation of ZB. In other words, we can directly observe the ZB by measuring the time dependence of transmission coefficient of photons around the Dirac point for 2D thick PC slabs.

The above calculations only focus on one kind of structure parameters. In fact, if we change the dielectric constant and radius of the cylinder, for example, $\varepsilon$ changes from 8 to 14 and $R$ from 0.25a to 0.4a, similar phenomena can be obtained. That is to say, observing such a phenomenon does not need critical conditions. In addition, our numerical simulations are based on the multiple-scattering method. The multiple-scattering method is a very efficient way of handling the scattering problem of a finite sample containing cylinders of circular cross sections, and it is capable of reproducing accurately the experimental transmission data [20]. Thus, we are sure that our results are reliable and observable. We are looking forward to experimentally observing this effect in the future.

In summary, we have demonstrated that the ZB of photons exists near the Dirac point in 2D PC. The superiority of the PC system is that this phenomenon can not only be found in different scaling structures with wide frequency regions, but also the difficulties of observing the ZB in the electron systems can be overcome. In addition, the phenomenon can be observed by measuring the time dependence of transmission coefficients through the PC slabs. This facilitates observation of the effect from experimental measurements. We have observed such a phenomenon by exact numerical simulations. Thus, a long-standing theoretical prediction has been demonstrated.

We wish to thank D. Z. Zhang, Z.Y. Li, and Z. Y. Liu for useful discussions. This work was supported by the National Natural Science Foundation of China (Grant No. 10674017) and the National Key Basic Research Special Foundation of China under Grant 2007CB613205. The project was also sponsored by NCET and RFDP.

## References

[1] E. Schrödinger, Sitzungsber. Preuss. Akad. Wiss., Phys. Math. Kl. 24, 418 (1930).

[2] H. Feschbach and F. Villars, Rev. Mod. Phys. 30, 24 (1958); A. O. Barut and A. J. Bracken, Phys. Rev. D 23, 2454 (1981); A. O. Barut and W. Thacker, Phys. Rev. D 31, 1386 (1985); K. Huang, Am. J. Phys. 20, 479 (1952).

[3] F. Cannata, L. Ferrari, and G. Russo, Solid State Commun. 74, 309 (1990); L. Ferrari and G. Russo, Phys. Rev. B 42, 7454 (1990); F. Cannata and L. Ferrari, Phys. Rev. B 44, 8599 (1991).

[4] W. Zawadzki, Phys. Rev. B 72, 085217 (2005); T. M. Rusin and W. Zawadzki, J. Phys.: C19, 136219 (2007)

[5] D. Lurié and S. Cremer, Physica (Amsterdam) 50, 224 (1970).

[6] J. Schliemann, D. Loss, and R. M. Westervelt, Phys. Rev. Lett. 94, 206801 (2005); Phys. Rev. B 73, 085323 (2006).

[7 ]S.-Q. Shen, Phys. Rev. Lett. 95, 187203 (2005).

[8] Z. F. Jiang, R. D. Li, S.-C. Zhang, and W. M. Liu, Phys. Rev. B 72, 045201 (2005).

[9] P. Brusheim and H. Q. Xu, Phys. Rev. B 74, 205307 (2006).

[10] Novoselov, K. S., Geim, A. K., Morozov, S. V., Jiang, D., Zhang, Y., Dubonos, S. V., Grigorieva, I. V., and Firsov, A. A., Science 306, 666 (2004).

[11] P. R. Wallace, Phys. Rev. 71, 622 (1947).

[12] For a review, see T. Ando, J. Phys. Soc. Jpn. 74, 777 (2005).

[13]M. I. Katsnelson, Eur. Phys. J. B 51, 157 _2006_.

[14]T. M. Rusin and W. Zawadzki, cond-mat/0702425 (unpublished).

[15] W. Zawadzki, Phys. Rev. B 74, 205439 (2006).

[16] J. Cserti and G. David, Phys. Rev. B 74, 172305 (2006).

[17] F. D. M. Haldane and S. Raghu, e-print arXiv:cond-mat/0503588; S. Raghu and F. D. M. Haldane, e-print arXiv:condmat/0602501.

[18] R. A. Sepkhanov, Ya. B. Bazaliy, and C. W. J. Beenakker, Phys. Rev. A **75**, 063813 (2007).

[19] X.D. Zhang, cond-mat/07100682.

[20] L.M. Li and Z.Q. Zhang, Phys. Rev. B **58**, 9587 (1998); X.D. Zhang, Z.Q. Zhang, L.M. Li, C. Jin, D. Zhang, B. Man, and B. Cheng, *ibid.* **61**, 1892 (2000).

[21] H. van Houten and C. W. J. Beenakker, in *Analogies in Optics and Micro Electronics*, edited by W. van Haeringen and D. Lenstra (Kluwer, Dordrecht, 1990).